\documentclass[prd,reprint]{revtex4-1}
\usepackage{slashed} 
\usepackage{caption,subcaption} 
\usepackage{nccmath} 
\usepackage{times}
\usepackage{amsmath} 
\usepackage{amssymb} 
\usepackage{verbatim} 
\usepackage{graphicx} 
\usepackage{color} 
\usepackage{booktabs} 
\usepackage{empheq}
\usepackage{relsize}
\usepackage{fancyref}
\usepackage{xcolor,colortbl}
\definecolor{olive}{rgb}{0.3, 0.4, .1}
\definecolor{fore}{RGB}{249,242,215}
\definecolor{back}{RGB}{51,51,51}
\definecolor{title}{RGB}{255,0,90}
\definecolor{dgreen}{rgb}{0.,0.6,0.}
\definecolor{gold}{rgb}{1.,0.84,0.}
\definecolor{JungleGreen}{cmyk}{0.99,0,0.52,0}
\definecolor{BlueGreen}{cmyk}{0.85,0,0.33,0}
\definecolor{RawSienna}{cmyk}{0,0.72,1,0.45}
\definecolor{Magenta}{cmyk}{0,1,0,0}
\definecolor{lcyan}{rgb}{0.6,1,1}
\usepackage[usenames,dvipsnames,pdf]{pstricks}
\usepackage{epsfig}
\usepackage{pst-grad} 
\usepackage{pst-plot} 
\usepackage{pstricks-add}
\usepackage[off]{auto-pst-pdf} 
 
\newcommand{\lp}{\left(} \newcommand{\rp}{\right)} 
\newcommand{\lb}{\left\{} \newcommand{\rb}{\right\}} 
\newcommand{\ls}{\left[}  \newcommand{\rs}{\right]}
\newcommand{\lv}{\left|}  \newcommand{\rv}{\right|}

\newcommand{\cd}{\!\cdot\!}
\newcommand{\st}[1]{\slashed{#1}}
\mathchardef\mhy="2D   

\begin{document}
\title{Exact solutions of the bound Dirac and Klein Gordon equations in non co propagating electromagnetic plane waves}
\author{A. Hartin}
\affiliation{University College London, Gower Street, London WC1E 6BT, UK}
\date{\today}
\maketitle
\section{Abstract}

A new class of exact solutions of the bound Dirac and bound Klein Gordon equations in non co propagating plane waves is found. The solutions are based on the physical principle of maintaining local gauge invariance in the Furry picture Lagrangian when N external fields can undergo independent gauge transformations. The solutions can be expressed in terms of the Hamilton Jacobi action and a gauge invariant effective particle momentum in the ensemble of external fields. Rotations of the effective particle momentum, which preserve local gauge invariance, are introduced into the action using matrix calculus. The set of such rotations provides the class of new solutions constituting a family of Volkov like solutions for one external field. When applied to two or more non co propagating external fields, the rotational symmetry provides counter terms which decouple the fields. The bound state equations of motion become solvable for any number of non co propagating external fields. Through angular spectral decomposition, which represents electromagnetic fields of any form as a spatial Fourier series of non co propagating plane waves, the new solutions described here can be applied to strong field physics problems in any external electromagnetic field.

\section{\label{sect:intro}Introduction}

Particle physics interactions that occur in the presence of ultra intense electromagnetic fields, present a range of novel phenomena that give new insights into the quantum vacuum \cite{DiPiaz12,EhlKam09,Hartin18a}. Experiments designed to study such strong field effects, often take place with a strong field provided by an intense laser. The calculation of strong field transition probabilities proceeds with certain assumptions about the structure of the laser field. \\

The most common assumption is that the laser field can be represented by a non depletable, plane wave electromagnetic field. Then, the field can be considered classical and be included in the strong field QED Lagrangian using the semi classical Furry picture \cite{Furry51,Schweber62,JauRoh76,Hartin11a}. The Furry picture results in the minimally coupled bound Dirac field and associated equation of motion. Exact solutions of the bound Dirac, as well as the bound Klein Gordon equations are an essential part of any strong field physics calculation. \\

The Volkov solution of the minimally coupled Dirac equation in a single plane wave is a long standing result \cite{Volkov35}. Various attempts have been made to extend the Volkov solution to other plane wave combinations, with \cite{Sengupta67} providing a solution for an external field consisting of two polarised plane electromagnetic waves. \cite{Bagrov74,Bagrov75} considered a relativistic electron interacting with a co propagating quantised and classical plane wave, while a fully quantised field can be treated by use of coherent states \cite{Glauber63,Glauber63b}.\\

Solutions of the bound Dirac equation exist also for Coulomb fields, longitudinal fields, pulsed fields, and fields of colliding charge bunches \cite{BagGit90,BocFlo09,HeiSeiKam10,HeiIldMar10,Hartin15,MacDiP11,Mackenroth14,SeiKam11,Seipt12,Seipt17}. \cite{Varro14} created a class of solutions for charged particles travelling in a medium, and pair creating electric fields were considered by \cite{GavGit17}. Real laser pulses are strongly focussed as well as pulsed. The associated electromagnetic fields are complicated and there have been attempts to take them into account \cite{Hartin91,Derlet95,Harvey16}. \\

Exact solutions in two plane waves are most easily obtained when they are co propagating. In that case, starting from a Volkov ansatz, the differential equations are separable and remain first order. The resultant solution is a simple function of Volkov actions for each of the component fields \cite{Lyulka75,Pardy06,NarFof96,BerVal73}. Solutions in two non co propagating fields, however, have been problematic except in some special cases. The intense, constant crossed fields associated with ultra relativistic charge bunches approaching collision, are nominally non co propagating. However in this case, Lorentz transformations allow the reappearance of co propagation \cite{Hartin15}. \\

In the anti co propagating case (head on propagation of the two fields) which can always be obtained from non co propagating fields via Lorentz transformations, the differential equations appear to include coupled terms. This coupling between the two fields inhibits separability, leading to second order differential equations whose solutions, where they exist, contain specific analytic functions. Under special conditions, for the Klein Gordon equation, these are Mathieu functions which display non physical parameter regions \cite{KingHu16}. \\

In this paper, by contrast, an alternative solution of the bound Dirac and bound Klein Gordon equations for the case of two anti co propagating plane wave fields, with no other pre conditions, is provided. The methodology developed to achieve this will prove robust enough to also solve the general case of an infinite series of non co propagating plane wave fields. This infinite series of plane waves, promises application to a general electromagnetic field using principles of Fourier optics \cite{Goodman05}. \\

Whereas Huygens and Fresnel analysed a field in terms of spherical waves emitted from sources on the wave front, the angular spectral representation decomposes any field into a spatial Fourier series of plane waves \cite{Clemmow96}. Focussed laser fields in particular can be represented using such a technique \cite{Capoglu13}. The general solution provided in this paper, in terms of just such a series of non co propagating plane waves, should therefore be applicable to theoretical calculations for strong field physics processes in all realistic external fields of interest. \\

The way to proceed in this paper is indicated by an earlier work on the calculation of Furry picture transition probabilities using Fierz relations. When an alternative form of the Volkov solutions can be cast into an alternative form, written in terms of the effective particle momentum $\Pi_\text{px}$, arising from the Hamilton Jacobi action and the external field 4-potential. When the alternative form of the Volkov solution is used, the analytic form of the transition probabilities became much simpler \cite{Hartin16}. \\

A closer examination of the effective strong field particle momentum, $\Pi_\text{px}$ reveals that it is a gauge invariant function with respect to the external field. This suggests a closer examination of the symmetries of the Furry picture Lagrangian, and that is indeed the starting point of this paper. \\

So, this paper intends to do the following: Local gauge invariance of the Furry picture Lagrangian will be examined with the aim of understanding the structure of the Volkov solution. In this way, the underlying physical role of the effective particle momentum in external fields will become clear. The class of exact solutions will be extended by allowing for rotations of the effective particle momentum. \\

The methodology that extends the class of solutions for one external field, will then be applied to the case of two non co propagating fields. The rotational symmetry allowed for the effective particle momentum, will provide the freedom to introduce counter terms into the solution ansatz. This will allow the removal of coupled terms and will provide the desired solution. Mathematically, the action in the new solution class will be reformulated using matrix calculus techniques. \\

In terms of conventions, a standard metric is adopted, natural units will be utilized, and the Lorenz gauge will be employed. The Lorenz gauge requires scalar products of the external field 4-momentum and 4-potential to vanish, $k.A^e=0$. This condition can obviously be extended, when $A^\text{e}$ is a series of component fields 

\begin{align}\label{eq:notn}
\text{metric: }&\quad g_{\mu\nu}=(1,-1,-1,-1) \notag\\
\text{units: }&\quad c=\hbar=4\pi\epsilon_0=1, \quad e=\sqrt{\alpha} \\
\text{gauge: }&\quad \partial A^\text{e}(k\cd x)=0 \implies k\cd A^\text{e}=0 \notag \\
& \quad \partial \sum_\text{i} A^\text{e}_\text{ix}(k\cd x)=0 \implies \sum_\text{i} k_\text{i}\cd A^\text{e}_\text{x}=0\notag 
\end{align}




To ease the notation, the charge e, accompanying the 4-potential will be implicitly assumed. So, the external field with 4-vector $k$, will be represented implicitly by $A^\text{e}_\text{x}\equiv eA^\text{e}(k\cd x)$. This paper deals with N external fields, whose 4-momenta are distinguished by subscripts $k_1,k_2,..,k_\text{N}.$. Where it is necessary to discuss components of these 4-vectors, they will be denoted as superscripts, $k_1\equiv(k_1^0,k_1^1,k_1^2,k_1^3)$. \\

Again to ease the notation, only positive energy solutions are written, though the analysis also delivers the negative energy solutions.


\section{\label{sect:localGI}Local gauge invariance in the Furry picture}

The stipulation of local gauge invariance in the QED Lagrangian, requires the Dirac field to couple to a massless gauge field $A_\mu$ whose gauge transformations cancel with those of the derivative of the Dirac field. The cancellation is achieved with the covariant derivative $D=i\partial-A_\text{x}$. Neglecting the Maxwell part,

\begin{gather}\label{eq1}
\mathcal{L}=\bar\psi (i\st{\partial}-\st{A}_\text{x}-m)\psi,\,\, \psi\!\rightarrow\! e^{-i\alpha}\psi,\;\; A_\text{x}\!\rightarrow\! A_\text{x}+\partial\alpha
\end{gather}

The Furry picture, in contrast, accommodates an external gauge field $A^\text{e}_\text{x}$, as well as the vacuum gauge field $A_\text{x}$. Local gauge invariance then requires a cancellation from three independent gauge transformations: those of the two gauge fields as well as the external field dependent, bound Dirac field $\psi^\text{FP}(A^\text{e}_\text{x})$,

\begin{align}
\text{invariant}\quad & \mathcal{L}^\text{FP}=\bar\psi^\text{FP}\lp i\st{\partial}-\st{A}^\text{e}_\text{x}-m\rp \psi^\text{FP}-\,\bar\psi^\text{FP}\st{A}\,\psi^\text{FP} \notag\\
\text{under }\quad & A_\text{x} \rightarrow A_\text{x} +\partial_\mu\alpha, \quad A^\text{e}_\text{x}\rightarrow A^\text{e}_\text{x}+\partial\alpha^\text{e} \\
& \psi^\text{FP}(A^\text{e}_\text{x})\rightarrow e^{-i\alpha}\psi^\text{FP}(A^\text{e}_\text{x}+\partial\alpha^\text{e}) \notag
\end{align}

Since gauge transformations of the vacuum gauge field and bound Dirac field cancel each other, gauge transformations of the external field must vanish amongst themselves. The necessary behaviour can be made apparent by considering the well known Volkov solution \cite{Hartin18a}, which proceeds from the Furry picture, for the bound Dirac equation with a single plane wave electromagnetic field (from now on the Furry picture will be assumed implicitly),

\begin{gather}
\psi_\text{prx}=\ls 1 - \mfrac{\st{A}^\text{e}_\text{x}\st{k}}{2k\cdot p}\rs\,u_{pr}\, e^{-i\ls p\cdot x+\medint\int^{k\cdot x}\!\mfrac{2A^\text{e}_\xi\cdot p-A^{\text{e}\,2}_\xi}{2k\cd p}\,\text{d}\xi\rs}
\end{gather}

The covariant derivative of the Volkov solution returns a gauge invariant canonical momentum (GICM), $\Pi_\text{px}$ which satisfies the requirements of local gauge invariance

\begin{gather}
\st{D}\,\psi_\text{prx}=\st{\Pi}_\text{px}\psi_\text{prx},\quad D \equiv i\partial- A^\text{e}_\text{x} \notag\\
 \Pi_\text{px}=p-A^\text{e}_\text{x}+k\,\mfrac{2A^\text{e}_\text{x}\cd p-A^{\text{e}\,2}_\text{x}}{2k\cd p}
\end{gather}



In the next section, the GICM will be considered as a fundamental 4-vector, in constructing solutions of the bound Dirac equation for one plane wave field.
 
\section{\label{sect:volkov}Alternative form for the Volkov solution}

The usual strategy for constructing solutions of the bound Dirac equation is outlined in section 32 of \cite{BerLifPit82}. Here, a modified procedure for one external field $A^\text{e}_\text{x}$ with 4-momentum $k$ is described, such that the role of the GICM is explicit. \\

The second order bound Dirac equation is formed by operating on the first order bound Dirac equation with a conjugate operator, which in turn operates on the solution $\Phi_\text{px}$

\begin{gather}
(\st{D}+m)(\st{D}-m)\Phi_\text{px}=\lp D^2-m^2-i\,\st{k}{\st{A}^\text{e}_\text{x}}'\rp\Phi_\text{px}=0 \notag\\ 
\text{where }\quad D=i\partial-A^\text{e}_\text{x},\quad {\st{A}^\text{e}_\text{x}}'\equiv \mfrac{\text{d}\,\st{A}^\text{e}_\text{x}}{\text{d}(k\cd x)} 
\label{eq:2ndordDE}\end{gather}

The scalar part of the above equation, which coincides with the Hamilton Jacobi equation, is solved in order the obtain the action. The GICM, $\Pi_\text{px}$ appears as the covariant derivative of the classical action,

\begin{gather}
(D^2\!-\!m^2)\, e^{-iS_\text{px}}=(\Pi_\text{px}^2\!-\!m^2)\, e^{-iS_\text{px} }=0 \\[4pt]
S_\text{px}=p\cd x+\medint\int^{k\cdot x} \mfrac{2A^\text{e}_\xi\cd p-A^{e 2}_\xi}{2k\cd p}\,\text{d}\xi, \quad De^{-iS_\text{px}}=\Pi_\text{px}\,e^{-iS_\text{px}} \notag
\end{gather}

The GICM has several useful properties in addition to its gauge invariance. The magnitude of the GICM lies on the free mass shell, the scalar product with it's space-time derivative vanishes, as does the doubled slashed derivative written in exponential form,

\begin{gather}\label{eq:gicmprops}
\Pi_\text{px}^2=m^2, \quad \partial\cd\Pi_\text{px}=0, \quad\st{\partial}\st{\Pi}_\text{px}\,e^{-\medint\int^{k\cdot x}\!\mfrac{\st{\partial}\,\st{\Pi}_{\text{p}\xi}}{2k\cd \Pi_{\text{p}\xi}}\,\text{d}\xi}=0
\end{gather}

The slashed derivative of the GICM returns the spinor part of the second order bound Dirac equation so that the positive energy solution can be written, 

\begin{gather}
\Phi_\text{px}=e^{- iS_\text{px} -\medint\int^{k\cdot x}\!\mfrac{\st{\partial}\,\st{\Pi}_{\text{p}\xi}}{2k\cd \Pi_{\text{p}\xi}}\,\text{d}\xi}\equiv e^{- iS_\text{px} +\mfrac{\st{k}\,\st{A}^\text{e}_\text{x}}{2k\cd\Pi_\text{px}}}
\end{gather}

According to the procedure of \cite{BerLifPit82}, the solution to the first order bound Dirac equation, $\psi_\text{prx}$ is then obtained by operating with the slashed covariant derivative on $\Phi_\text{px}$. A four component spinor $u_\text{0r}$ for a fermion of spin r at rest, is also included in order to recover the free fermion solution when the external field is not present,

\begin{align}
\psi_\text{prx}&=(\st{D}+m)\Phi_\text{px}\,u_\text{0r} \notag\\
&=\ls \st{\Pi}_{\text{px}}+ m \rs \, e^{- iS_\text{px} +\mfrac{\st{k}\,\st{A}^\text{e}_\text{x}}{2k\cd\Pi_\text{px}}}\,u_\text{0r} \\[4pt]
&\rightarrow \ls \st{p}+ m \rs u_\text{0r}\,e^{-ip\cdot x} = u_\text{pr}\,e^{-ip\cdot x} \;\text{ as }\; \lv A^\text{e}_\text{x}\rv\rightarrow 0\notag
\end{align}

This so called alternative form for the Volkov solution, which proves useful for the efficient calculation of Furry picture transition probabilities \cite{Hartin16} can be transformed into the standard Volkov form by manipulating the spinor,

\begin{gather}
\ls \st{\Pi}_{\text{px}}+ m \rs e^{\mfrac{\st{k}\,\st{A}^\text{e}_\text{x}}{2k\cd\Pi_\text{px}}}\,u_\text{0r}=\ls \st{\Pi}_{\text{px}}+ m \rs \ls 1+ \mfrac{\st{k}\st{A}^\text{e}_\text{x}}{2k\cd p}\rs u_\text{0r} \notag\\
=\ls 1+ \mfrac{\st{k}\st{A}^\text{e}_\text{x}}{2k\cd p}\rs \ls \st{p}+ m\rs u_{0r}=\ls 1+ \mfrac{\st{k}\st{A}^\text{e}_\text{x}}{2k\cd p}\rs u_\text{pr} \notag\\
=\ls \st{\Pi}_{\text{px}}+ m \rs \mfrac{\st{k}}{2k\cd p} u_\text{pr} 
\end{gather}

The purpose of deriving the Volkov solution in this way was to establish a procedure to obtain the solution expressed in terms of the GICM,  $\Pi_\text{px}$. The GICM preserves local gauge invariance and is the effective particle momentum for charged particles within strong external fields. By finding operations which preserve the properties of $\Pi_\text{px}$, a whole family of Volkov like solutions can be found.

\section{Families of Volkov-like solutions}

Now, a procedure for finding families of solutions for the case of one external plane wave field will be developed. The gauge invariance of the GICM $\Pi_\text{px}$ will be considered essential to it's physical role. Extensions to the GICM, obtained by including extra terms which preserve it's properties, will be sought. This will be the ansatz by which a new class of solutions to the bound Dirac and bound Klein Gordon equation will be found. \\

The desired extension to a family of GICMs in one external field $\Pi^\text{1F}_\text{px}$, include an extra term consisting of a 4 vector $n$ and function $f$ of the scalar product $k\cd x$. Requiring this family of GICMs to remain on the mass shell and to vanish when dotted with the derivative operator, is sufficient to determine the form of $n$ and $f$,

\begin{gather}
\Pi^{1\text{F}}_\text{px}= \pi_\text{px}+n f(k\cd x),\quad \pi_\text{px}\equiv p+\sigma_\text{px} \\
\sigma_\text{px}\equiv k \Omega_\text{px}-A^\text{e}_\text{x}, \quad \Omega_\text{px}=k\,\mfrac{2A^\text{e}_\text{x}\cd p-A^{e 2}_\text{x}}{2k\cd p} \notag \\
\Pi_{\text{px}}^\text{1F 2}\!=\!m^2,\, \partial\cd  \Pi_{\text{px}\mu}^\text{1F}\!=\!0\! \implies\!n\cd k\!=\!0, f\!=\!\ls 0,-\mfrac{2\,n\cd \pi_\text{px}}{n^2}\rs \notag
\end{gather}

These conditions restrict the 4-vector $n$ to being orthogonal to the external field 4-momentum $k$. The function of scalar products $f$, is a set depending on the remaining freedom for the choice of $n$. The null member $f=0$ is simply the GICM without the additional term, i.e. it corresponds to the Volkov solution. The second part of the set $f=-\tfrac{2n\cdot \pi_\text{px}}{n^2}$, together with the 4-vector $n$ constitutes the bulk of the family of GICMs. Since $f$ is a function of the gauge invariant part of the GICM $\sigma_\text{px}$, gauge invariance is maintained. \\

Next, the family of GICMs should appear when the covariant derivative operator in the bound Dirac and Klein Gordon equations operates on them. So, the action has also to be extended to a family of actions,

\begin{gather}
De^{-iS^\text{1F}_\text{px}}=\Pi^\text{1F}_\text{px}\,e^{-iS_\text{px}}
\end{gather}

The family of actions $S^\text{1F}_\text{px}$ has to deliver the term $nf(k\cd x)$. In order to do this, a mathematical application from matrix calculus can be employed (see appendix \ref{app:matcalc}). A rotation matrix $R$ and its inverse $R^{-1}$ can be applied to a scalar product, if it is decomposed into basis vectors $e^i$ and vector components $n^\text{i},k^\text{i},x^\text{i}$. The $\;\widetilde{ }\;$ notation is used to denote the decomposition

\begin{gather}
R\,\widetilde{n\cd x}\equiv k\cd x,\quad R^{\,\mhy1}\widetilde{k\cd x}\equiv n\cd x \\
\widetilde{n\cd x}\equiv \sum^4_{i=1} e^i n^\text{i} x^\text{i}, \quad e^i=\lb \begin{pmatrix} 1 \\ 0 \\ 0 \\ 0 \end{pmatrix},\begin{pmatrix} 0 \\ 1 \\ 0 \\ 0 \end{pmatrix},\begin{pmatrix} 0 \\ 0 \\ 1 \\ 0 \end{pmatrix},\begin{pmatrix} 0 \\ 0 \\ 0 \\ 1 \end{pmatrix} \rb
 \notag
\end{gather}

The rotation matrix is deployed in the family of actions, requiring an integration consisting of matrix limits, matrix integrand and matrix measure \cite{Anderson03,Gentle17}, and the required result is obtained,

\begin{gather}
S_\text{px}^{1\text{F}}=p\cd x+\medint\int^{k\cdot x}\!\Omega_{\text{p}\xi}\,\text{d}\xi+\medint\int^{\widetilde{n\cdot x}}\!f(R\widetilde{\xi})\,\text{d}\widetilde{\xi} \\
\Pi^\text{1F}_\text{px}=p-A^\text{e}_\text{x}+k\Omega_{px}+nf(k\cd x) \notag
\end{gather}

The rotation is permitted in the action as long as none of the theory's invariance principles are violated. The action continues to satisfy the equations of motion for a fermion embedded in an external field $A^\text{e}$ with a particular 4-momentum $k$. This rotation R, should not be confused with a Lorentz transformation since it does not apply to the coordinate system in general. \\

The family of actions is sufficient to write down a family of solutions $\phi^\text{1F}_\text{px}$ for the bound Klein Gordon equation. For the bound Dirac equation, an additional spinor is required. For the Volkov solutions, the product of the GCIM and the additional spinor vanishes when operated on by the slashed derivative due to the properties of the GICM. This property also holds for the family of GICMs

\begin{gather}
i\st{\partial}_\text{x}\st{\Pi}^\text{1F}_\text{px} \;e^{-\medint\int^{k\cdot x}\!\mfrac{\st{\partial}_\xi\st{\Pi}^\text{1F}_{\text{p}\xi}}{2k\cd\Pi^\text{1F}_{\text{p}\xi}}\,\text{d}\xi}=0
\end{gather}

So, a family of solutions $\phi^\text{1F}_\text{px},\psi^\text{1F}_\text{prx}$ for both the bound Klein Gordon and bound Dirac equations can be written in terms of the family of GICMs, 

\begin{gather}
(D^2-m^2)\,\phi^{1\text{F}}_{\text{px}}=0,\quad  (\st{D}-m)\,\psi^{1\text{F}}_{\text{prx}}=0\\
\phi^{1\text{F}}_{\text{px}}=e^{-iS_\text{px}^{1\text{F}}},\quad\psi^{1\text{F}}_{\text{prx}}\!=\!\ls \st{\Pi}^{\text{1F}}_{\text{px}} + m \rs\! e^{-iS_\text{px}^{1\text{F}}-\medint\int^{k\cdot x}\!\mfrac{\st{\partial}_\xi\st{\Pi}^\text{1F}_{\text{p}\xi}}{2k\cd\Pi^\text{1F}_{\text{p}\xi}}\,\text{d}\xi }\,u_\text{0r} \notag
\end{gather}

To elucidate what has been done in this section, a particular example is given. A simple configuration where an electron $p$ collides head-on with a circularly polarised external electromagnetic field $A^\text{e}_\text{x},\, k\equiv(w,0,0,w)$, is considered. Choosing for the family of solutions a space-like unit 4-vector $n$, the 3-vector part of $n$ must lie in the plane defined by the external field 3-potential,

\begin{gather}
p\equiv(\varepsilon_\text{p},0,0,p^\text{3}), \, A^\text{e}_\text{x}\equiv (0,|A^\text{e}|\cos k\cd x,|A^\text{e}|\sin k\cd x,0),\,  \notag\\
k\equiv(w,0,0,w),\;\;n_\mu\equiv(0,\vec{n}_\perp,0),\;\; |\vec{n}_\perp|=1, \, n^2=-1
\end{gather}

With such an arrangement, several of the scalar products vanish and the analytic expressions simplify. There still remains a subset of the family of solutions, given by the freedom for the vector $\vec{n}_\perp$ to rotate azimuthally around the z-axis, within the plane defined by the 3-potential. \\



The explicit form of the rotation $R\widetilde{\xi}$ is introduced via the matrix calculus relations of Appendix \ref{app:matcalc}, and the family of Volkov-like actions, associated GICMs and additional spinor are,

\begin{gather}
S^\text{1F}_\text{px}=p\cd x-\medint\int^{k\cdot x}\!\mfrac{A^{\text{e}\,2}_\xi}{2k\cd p}\,\text{d}\xi-2\medint\int^{\widetilde{n\cdot x}}\!n\cd A^\text{e}_{R\widetilde{\xi}}\,\text{d}\widetilde{\xi} \notag\\
\Pi^\text{1F}_\text{px}=p-A^\text{e}_\text{x}-k\mfrac{A^{\text{e}\,2}_\text{x}}{2k\cd p}-n\, 2n\cd A^\text{e}_\text{x} \\
\mfrac{\st{\partial}_\xi\st{\Pi}^\text{1F}_{\text{p}\xi}}{2k\cd\Pi^\text{1F}_{\text{p}\xi}}=\mfrac{\st{k}\lp\st{A}'_\xi-\st{n}\,2n\cd A'_\xi\rp}{2k\cd p} \notag
\end{gather}

This methodology employed to produce the family of solutions, can be used for more complicated external field combinations. For two or more non co directional external fields, traditional methods lead to coupled terms which are difficult to handle. However, now there is flexibility to look within the set of rotated solutions for additional terms that provide decoupling.

\section{Dirac and Klein Gordon equation solutions in non co directional external plane wave fields}

Solutions to the bound Dirac and bound Klein Gordon equations are now sought for a sum of non propagating plane waves. Such a sum can serve as a Fourier decomposition which can be used to model any form of phase front by the technique of Fourier optics \cite{Goodman05}. The external field 4-potential is written as a sum of terms,

\begin{gather}
 A^\text{e}_\text{x}=\sum\limits_{i=1}^N A^\text{e}_\text{ix}(k_\text{i}\cd x),\,\, k_\text{i}^2=0,\,\, k_\text{i}\cd k_\text{j}\neq 0\; \forall\; i\neq j  
\end{gather}



The case of two ($N=2$) non co propagating plane wave external fields, is dealt with first. The techniques developed to deal with this case will then be extended to multiple non co propagating fields. \\

With two external fields, there must be gauge invariance with respect to both fields independently, so it seems right to allow two independent rotations, $R_\text{1},R_\text{2}$ to appear in the action. The rotation matrices will be applied to a general 4-vector $n_\text{12}$ in order to rotate it onto the 4 momenta of the external fields, $k_\text{1},k_\text{2}$.  A general function $f_\text{12}$ of the two scalar products ($k_\text{1}\cd x$, $k_\text{2}\cd x$) will also be introduced. So, the provisional family of actions and associated GICMs in the two external fields can be written,

\begin{gather}
S_\text{px}^{2\text{F}}=p\cd x+\sum_{i=1}^2\medint\int^{k_\text{i}\cdot x}\!\Omega_{\text{ip}\xi}\,\text{d}\xi-\medint\int^{\widetilde{n_\text{12}\cdot x}}\!f_\text{12}(R_1\widetilde{\xi},R_2\widetilde{\xi})\,\text{d}\widetilde{\xi} \notag\\
\Pi_\text{px}^{2\text{F}}=\pi_\text{12}-n_{\text{12}}\,f_\text{12}(k_1\cd x,k_2\cd x), \quad \pi_\text{12}\!\equiv\! p\!+\!\sigma_\text{1}\!+\!\sigma_\text{2} \\
\sigma_\text{i}\equiv k_\text{i} \,\Omega_\text{ipx}-A^\text{e}_\text{ix},\quad \Omega_\text{ipx}\equiv\mfrac{2A^\text{e}_\text{ix}\cd p-A^{\text{e}\,2}_\text{ix}}{2k_\text{i}\cd \Pi^\text{2F}_\text{px}} \notag
\end{gather}

To satisfy the desired properties of the GICM, the 4-vector $n_\text{12}$ should be orthogonal to the 4-momenta of both the external fields, so that again $\partial\cd\Pi^\text{2F}_\text{px}=0$. This could be achieved, for example, by setting the zeroth component of $n_\text{12}$ to zero, and the 3-vector part orthogonal to the plane formed by $(\vec{k}_\text{1},\vec{k}_\text{2})$. \\

A more general 4-vector $n_\text{12}$, orthogonal to light like 4-vectors $k_1$ and $k_2$, can be constructed with the aid of any other 4-vector $c$. This result is non trivial ($n_{12}\neq0$) as long as $c$ is not a linear combination of the original light like vectors,

\begin{gather}
n_{12}=c-\mfrac{k_{1}\,k_2\cd c+k_{2}\,k_1\cd c}{k_1\cd k_{2}},\quad n_\text{12}\cd k_1\!=\!n_\text{12}\cd k_2\!=\!0
\end{gather}

As in the one external field case, the unknown function $f_\text{12}$ is solved for by requiring the GICM to be on-shell,

\begin{gather}
\Pi^{\text{2F}}_\text{px}\!=\pi_\text{12}+\!n_{12}\,f_\text{12}  \\
\ls \Pi^{\text{2F}}_{\text{px}}\rs^2\!=\!m^2\!\implies\! f_\text{12}=-\mfrac{n_\text{12}\cdot\pi_\text{12}}{n_\text{12}^{\;\;2}}\pm\sqrt{\ls \tfrac{n_\text{12}\cdot\pi_\text{12}}{n_\text{12}^{\;\;2}}\rs^2 \!-\!\tfrac{2\sigma_\text{1}\cdot\sigma_\text{2}}{n_\text{12}^{\;\;2}}} \notag
\end{gather}

The extra term $n_\text{12}f_\text{12}$ is in effect, a counter term which cancels the term which couples the two fields $\sigma_\text{1}\cd \sigma_\text{2}$, in the square of the GICM. The particular solution for the quadratic equation in $f_\text{12}$ is constrained by the condition that it must vanish when the two external fields decouple, i.e. $f_\text{12}=0 \text{ when }\sigma_\text{1}\cd\sigma_\text{2}=0$. Since $f_\text{12}$ is a function of the gauge invariant $\sigma_\text{1}$ and $\sigma_\text{2}$, gauge invariance in the family of two field GICMs is maintained. \\

Now in the bound Dirac equation solution, only the additional spinor is required. Since the GICM contains coupled terms (they are only cancel when the square is taken), matrix calculus has to be deployed again. It is desireable to have the extra spinor together with the slashed GICM vanish under the slashed derivative, as before. That is achievable with,

\begin{gather}
i \st{\partial}_\text{x} \lp\st{\Pi}^\text{2F}_\text{px} \,\exp\!\ls-\sum_{i=1}^2\!\medint\int^{\widetilde{k_i\cdot x}}\! \st{\Delta}^\text{2F}_{\text{ip}\widetilde{\xi}}\,\text{d}\widetilde{\xi}\rs\rp=0 \\
\st{\Delta}^\text{2F}_{\text{ip}\widetilde{\xi}}\equiv\mfrac{\st{k}_i\lp\st{A}^{\text{e}'}_{i\widetilde{\xi}}-\st{n}_\text{12}\,2n_\text{12}\cd A^{\text{e}'}_{i\widetilde{\xi}}\rp}{2k_i\cd \Pi^\text{2F}_{\text{p}\widetilde{\xi}}}\notag
\end{gather}

Since the additional spinor is a function of the 4-momenta of both fields, a rotation that converts one into the other is required. So, the first component of the additional spinor contains a rotation $R_\text{12}$ that converts $k_1$ into $k_2$, and vice versa for the second component,

\begin{gather}
\st{\partial}_\text{x}\medint\int^{\widetilde{k_1\cdot x}}\st{\Delta}^\text{2F}_{\text{1p}\widetilde{\xi}}(\widetilde{\xi},R_\text{12}\widetilde{\xi})\,\text{d}\widetilde{\xi}=\st{k}_1\st{\Delta}^\text{2F}_\text{1px}(k_1\cd x,k_2\cd x) \\
\st{\partial}_\text{x}\medint\int^{\widetilde{k_2\cdot x}}\st{\Delta}^\text{2F}_{\text{2p}\widetilde{\xi}}(R_\text{21}\widetilde{\xi},\widetilde{\xi})\,\text{d}\widetilde{\xi}=\st{k}_2\st{\Delta}^\text{2F}_\text{2px}(k_1\cd x,k_2\cd x) \notag
\end{gather}

All the items are now in place, and the bound Dirac and bound Klein Gordon equation solutions in two non co-directional fields can be written with the same form as in the one external field case,

\begin{gather}\label{eq:2Fsol}
\phi^{2\text{F}}_{\text{px}}=e^{-iS_\text{px}^{2\text{F}}},\;\;\psi^{2\text{F}}_{\text{prx}}\!=\!\ls \st{\Pi}^{\text{2F}}_{\text{px}} + m \rs\! e^{-iS_\text{px}^{2\text{F}}-\sum\limits_{i=1}^2\int^{\widetilde{k_i\cdot x}}\! \st{\Delta}^\text{2F}_{\text{ip}\widetilde{\xi}}\,\text{d}\widetilde{\xi}}\,u_\text{0r}
\end{gather}

To continue to the general case of N non co propagating external field components, only a 4-vector orthogonal to the N 4-momenta, $k_1,k_2,...,k_\text{N}$ is needed. In the case of N=3, the required vector can be constructed from orthogonal 2-vectors $n_\text{12},n_\text{13},n_\text{23}$, and general vector $c$, where $c$ is not a linear combination of $k_\text{1},k_\text{2},k_\text{3}$,

\begin{gather}
n_{123}=c-n_{12}\mfrac{k_3\cd c}{k_3\cd n_{12}}-n_{13}\mfrac{k_2\cd c}{k_2\cd n_{13}}-n_{23}\mfrac{k_1\cd c}{k_1\cd n_{23}}
\end{gather}

To proceed to the orthogonal N-vector ($n_\text{N}\equiv n_\text{1..N}$) the above construction is simply iterated. The function $f_\text{N}$ which provides the decoupling terms appears once again from a quadratic equation, and contains all coupled gauge invariant scalar product combinations $\sigma_\text{i}\cd \sigma_\text{j}$. The extra spinor is additively separable under slashed differentiation, and its form is easily determined. So, the solutions in N non co propagating fields are,

\begin{gather}\label{eq:fullsol}
\phi^{N\text{F}}_{\text{px}}=e^{-iS_\text{px}^{N\text{F}}},\;\;\psi^{N\text{F}}_{\text{px}}\!=\!\ls \st{\Pi}^{\text{NF}}_{\text{px}} + m \rs\! e^{-iS_\text{px}^{N\text{F}}-\sum\limits_{i=1}^N\int^{\widetilde{k_i\cdot x}}\! \st{\Delta}^\text{NF}_{\text{ip}\widetilde{\xi}}\,\text{d}\widetilde{\xi}}\,u_\text{0r} \notag\\
\Pi^\text{NF}_\text{px}=\pi_\text{N}+n_\text{N}f_\text{N}, \quad \pi_\text{N}=p+{\textstyle \sum\limits_{i=1}^\text{N}} \sigma_\text{i},\quad \sigma_\text{i}=k_\text{i}\Omega_\text{ipx}-A^\text{e}_\text{ix} \notag\\
f_\text{N}=-\mfrac{n_\text{N}\cdot\pi_\text{N}}{n_\text{N}^{\;\;2}}\pm\sqrt{\ls \tfrac{n_\text{N}\cdot\pi_\text{N}}{n_\text{N}^{\;\;2}}\rs^2 \!-\!{\textstyle \sum\limits_{i,j=1,i\neq j}^\text{N}}\mfrac{2\sigma_\text{i}\cdot\sigma_\text{j}}{n_\text{N}^{\;\;2}}} \\
S_\text{px}^{N\text{F}}=p\cd x+{\textstyle \sum\limits_{i=1}^\text{N}}\medint\int^{k_\text{i}\cdot x}\!\Omega_{\text{ip}\xi}\,\text{d}\xi-\medint\int^{\widetilde{n_\text{N}\cdot x}}\!f_\text{N}(R_1\widetilde{\xi},..,R_N\widetilde{\xi})\,\text{d}\widetilde{\xi} \notag
\end{gather}

As before, the N rotations $R_1,..,R_N$, convert $n_\text{N}$ to each of the external field 4-momenta in turn. The form of the general solutions in an infinite series of non co propagating fields is notable, in that they are not much more involved than the two non co propagating fields solution. This offers the prospect of realistic Furry picture transition probability calculations in general external fields. For now, in the next section, a specific case for $N=2$ is examined.

\section{Solutions in two anti co propagating fields}

Bound Dirac and bound Klein Gordon equation solutions in two external fields have been attempted by several authors \cite{Pardy06,KingHu16}. In the case where the two fields are not co propagating, Lorentz transformations allow the problem to be considered in the anti co propagating configuration

\begin{gather}
A^\text{e}_\text{x}=A^\text{e}_\text{1x}(k_\text{1}\cd x)+A^\text{e}_\text{2x}(k_\text{2}\cd x) \\
 k_\text{1}\cd A^\text{e}_\text{1x}=k_\text{2}\cd A^\text{e}_\text{2x}=k_\text{1}\cd A^\text{e}_\text{2x}=k_\text{2}\cd A^\text{e}_\text{1x}=0, \quad k_\text{1}\cd k_\text{2}\neq 0 \notag
\end{gather}

With the $\vec{n}_\text{12}$ 3-vector being chosen perpendicular to the line of the counter propagating external field 3-momenta, the zeroth component must be zero and a unit magnitude can be chosen, $n_\text{12}\equiv(0,\vec{n}_\text{12}), \, n_\text{12}^2=-1$. There is still an azimuthal freedom for $\vec{n}_\text{12}$ within the plane formed by the two external field 3-potentials $\vec{A}^\text{e}_\text{1x},\vec{A}^\text{e}_\text{2x}$. With these specifications, the GICM of the new solutions is,

\begin{gather} 
\Pi^\text{2F}_\text{px}\!=\!\pi_\text{12}\!+\!n_\text{12}\!\ls n_\text{12}\cd (p\,\mhy A^\text{e}_\text{x})\!-\!\sqrt{\ls n_\text{12}\cd (p\,\mhy A^\text{e}_\text{x})\rs^2+2\sigma_1\cd \sigma_2}\rs \notag\\
\sigma_1\cd \sigma_2=k_1\cd k_2 \,\Omega_\text{1px} \,\Omega_\text{2px}+A^\text{e}_\text{1x}\cd A^\text{e}_\text{2x}
\end{gather} 

If the condition for zero transverse canonical momentum holds, the GICM simplifies further,

\begin{gather}\label{eq:ztcm} 
\Pi^\text{2F}_\text{px}\!=\!p\!+\!\sigma_1\!+\!\sigma_2\!-\!n_\text{12}\sqrt{2\sigma_1\cd \sigma_2}, \quad (p\,\mhy A^\text{e}_\text{px})_\perp\!=\!0 
\end{gather}

Unlike for a single external field, the family of solutions in two fields, expressed by the vector $n_\text{12}$ and its coefficient function, $f_\text{12}$, does not contain the null $f_\text{12}=0$ case. That is, there is no situation when there is not a counter term to provide decoupling of the two fields within the GICM. With that in mind, a comparison can be made to the bound Klein Gordon equation solutions in anti co propagating external fields studied by \cite{KingHu16}. In this earlier work, two subsets of the anti co propagating configuration were considered, the dynamics at a magnetic mode and the case of zero transverse canonical momentum. \\

\cite{KingHu16} used a Volkov ansatz for the initial trial solution, and the resulting second order Mathieu equation lead to solutions written in terms of Mathieu characteristic exponents. The behaviour of the Mathieu exponents is such that the solutions became unphysical across regions of parameter space. The only point where the solution remained consistently physical, was when the field strengths of both fields vanished. \\

By contrast, the GICM based solutions developed here, are physical across all parameter ranges. The reason for the contrasting results compared to those of \cite{KingHu16} is that the ansatz is different. By extending the class of possible solutions to a family of solutions based on the physical role of satisfying local gauge invariance, decoupling counter terms were automatically introduced. So, the solution set for anti co propagating fields found here, is mutually exclusive to those of \cite{KingHu16}, except for the special decoupling limit of vanishing external field strength. \\

\section{Conclusion}

Particle physics processes in the presence of intense electromagnetic fields, and the theoretical tools used to describe them, promise to give us a new way to test, among other things, the behaviour of the quantum vacuum under extreme conditions. Often, the experiments that test this strong field regime, employ intense electromagnetic fields that cannot be described by a single plane wave. In contrast, many long standing strong field analyses, including most of the strong field transition probability calculations, rely on solutions of the bound Dirac equation in just such an external field. \\

The work performed in this paper, to develop a new class of exact solutions of the Dirac and Klein Gordon equations in the presence of external fields, is intended to provide further theoretical tools for analysis of strong field experiments. As examples of applications of these new solutions, two experimental tests can be envisaged. \\

Two or more intense lasers can be brought to bear on an interaction region in which strong field processes take place. In this case, there is clearly the need to account for non co propagating fields and to explore in detail the angular dependence as well as the relative laser frequency and intensity effects on the strong field processes. In fusion projects which seek to initiate nuclear reactions via implosion from multiple strong lasers, the analysis here would also be of benefit. \\
 
In any case, very intense lasers are strongly focussed, and the electromagnetic field at the focal point is anything but simple. Efforts have been made in the past to account for such fields in strong field processes, but the full description has been difficult \cite{Hartin91,Derlet95,Harvey16}. The work here provides a new way to tackle this problem by using a Fourier decomposition into a plane wave series. \\

Previous theoretical studies have found it difficult to find a description for one of the simplest multiple plane wave configurations, that of two anti co propagating fields. Usually, a Volkov ansatz is used to develop a model solution. Once inserted in the bound Dirac and Klein Gordon equations, second order differential equations result which, even in restricted cases, lead to solutions with analytic functions that display non physical parameter regions \cite{KingHu16}. \\

A different ansatz was employed here, by examining first the Furry picture Lagrangian and the role played by the external field in maintaining overall local gauge invariance. The quantised gauge boson field already absorbs local gauge transformations of the bound Dirac field, so the external field contained within the bound Dirac field must be self invariant with respect to its own gauge transformations. \\

For a single plane wave external field, the Volkov solution of the bound Dirac equation provides exactly this gauge invariance. The relevant quantity, referred to here as the gauge invariant canonical momentum (GICM) is based on the action of the fermion in the external field. The GICM, $\Pi_\text{px}$ consists of the canonical momentum, $p-A^\text{e}_\text{x}$ and an additional term $k\Omega_\text{px}$ which absorbs local gauge transformations of the external field $A^\text{e}_\text{x}$. \\

The GICM seems to be a fundamental quantity in strong field physics interactions. Furry picture transition probabilities are simplified when expressed in terms of the GICM \cite{Hartin16}. Apart from gauge invariance, the GICM also lies on the free mass shell, $\Pi_\text{px}^2\!=\!p^2\!=\!m^2$ and it's scalar product with the space-time derivative vanishes, $\partial\cd\Pi_\text{px}=0$. \\

Under 4-rotations, the GICM retains it's mathematical properties and physical role. The set of 4-rotations was used to construct new families of solutions of the bound state equations. A mathematical mechanism for providing a 4-rotation was applied in the action of the solution. Using techniques of matrix calculus from the field of statistical analysis, scalar products were decomposed into a sum of basis matrices. Integrals of scalar products were shown to be equivalent to integrals containing matrix limits, matrix integrand and matrix measure. \\

The family of solutions based on rotated GICMs included solutions for the case of two anti co propagating fields. Problematic coupled terms were removed with counter terms, which led to first order equations and relatively simple, exact solutions of the bound Dirac and Klein Gordon equations. The methodology proved versatile enough for it to be extended to the most general case of N non co propagating external fields. \\

The new solutions found here, lead on to further work. It is essential to their purpose to apply these solutions to new calculations of Furry picture transition probabilities in new field configurations. To that end, the solutions need to be quantised, and as a prerequisite for that, their orthogonality and completeness have to be demonstrated. Additionally, spin sums and external field propagators in N non co propagating fields have also to be determined, This ongoing work is needed in order to carry out a general investigation of Furry picture transition probabilities in non co propagating fields.

\begin{acknowledgments}
The author would like to acknowledge funding from the Partnership of DESY and Hamburg University (PIER) for the seed project, PIF-2016-53. Additionally, this work was supported by a Leverhulme Trust Research Project Grant RPG-2017-143 and by STFC, United Kingdom.
\end{acknowledgments}

\appendix\section{Matrix calculus with rotation matrices}\label{app:matcalc}

A common operation for solutions of the Dirac equation in external fields, is the application of the partial derivative operator on the action such that,

\begin{gather}\label{eq:app1}
\mfrac{\partial}{\partial x^\mu} \medint\int^{k\cdot x}f(\xi) \,\text{d}\xi=k^\mu\, f(k\cd x)
\end{gather}

A relation that produces the same right hand side as equation \ref{eq:app1} can also be obtained, by rendering quantities into matrices and using matrix calculus. For instance, defining an integral $h$ that contains matrix limits, integrand and measure \cite{Gentle17,Anderson03}, vector components $k^\text{i},x^\text{i}$, and working first in two dimensions to illustrate the procedure,

\begin{gather}
 h\equiv[1\,\,1]\medint\int^{\tiny\begin{bmatrix}1\\0\end{bmatrix}k^1 x^1+\begin{bmatrix}0\\1\end{bmatrix}k^2 x^2} f\!\ls\left([1\,\, 0]+[0\,\, 1]\right)\xi\rs \text{d}\xi \notag\\
\mfrac{\partial}{\partial x^i} h=k^\text{i}\, f(k^1x^1+k^2x^2), \quad x^\text{i}\equiv(x^1,x^2)
\end{gather}

Then, in four dimensions, using four orthogonal basis row $e_i$ and column $e^i$ matrices that span the 4D domain of space-time, and defining a $\,\widetilde{ }\,$ notation for the vector decomposition of scalar product components, $k^\text{i},x^\text{i}$

\begin{gather}
\widetilde{k\cd x}\equiv \sum_{i=1}^4 e^i k^i x^i, \quad e^i\!=\!\lb\! \begin{pmatrix} 1 \\ 0 \\ 0 \\ 0 \end{pmatrix}\!,\!\begin{pmatrix} 0 \\ 1 \\ 0 \\ 0 \end{pmatrix}\!,\!\begin{pmatrix} 0 \\ 0 \\ 1 \\ 0 \end{pmatrix}\!,\!\begin{pmatrix} 0 \\ 0 \\ 0 \\ 1 \end{pmatrix} \!\rb
\end{gather}

With the aid of this decomposition, the matrix algebra equivalent of equation \ref{eq:app1}, implicitly including the initial unit row matrix $I_4\equiv[1\,1\,1\,1]$, is

\begin{gather}
\mfrac{\partial}{\partial x^\mu} \int^{\widetilde{k\cdot x}}f(\widetilde{\xi}) \,\text{d}\widetilde{\xi}=k^\mu\, f(k\cd x) \\
\medint\int^{\widetilde{k\cdot x}}f(\widetilde{\xi}) \,\text{d}\widetilde{\xi}\equiv I_4 \notag\medint\int^{\sum\limits_{i=1}^4 e^i k^i x^i}f\!\lp{\textstyle\sum\limits_{j=1}^4} e_j \xi\rp\,\text{d}\xi
\end{gather}

The advantage of introducing the matrix decomposition in this form, is that a rotation matrix $R$ that transforms the components of one vector into another, say $n\rightarrow k$ can be introduced,

\begin{gather}
\mfrac{\partial}{\partial x^\mu} \!\medint\int^{\widetilde{n\cdot x}}\!f(R\widetilde{\xi}) \,\text{d}\widetilde{\xi}=n^\mu\, f(k\cd x),\quad  R\widetilde{\xi}\equiv I_4 R\, \widetilde{\xi} \notag\\
R\widetilde{n\cd x}\equiv I_4 R{\textstyle \,\sum\limits_i}e^in^\text{i}x^\text{i}\equiv k\cd x 
\end{gather}

Due to the properties of rotation matrices, it is possible to use the inverse rotation $R^{-1}$ in a change of integration variable,

\begin{gather}
\medint\int^{\widetilde{n\cdot x}}\!f(R\widetilde{\xi}) \,\text{d}\widetilde{\xi}\equiv R^{-1}\!\medint\int^{\widetilde{k\cdot x}}\!f(\widetilde{\xi}) \,\text{d}\widetilde{\xi} \\
RR^{-1}=1, \quad R\widetilde{n\cd x}=k\cd x,\quad R^{-1}\widetilde{k\cd x}=n\cd x\notag 
\end{gather}

Since the space-time derivative operator can be brought inside the matrices, the rotation combined with the change of integration variable means that the space-time ($x$) dependence can be arbitrarily changed,

\begin{gather}
\mfrac{\partial}{\partial x} \medint\int^{\widetilde{n\cdot x}}\!f(R\widetilde{\xi}) \,\text{d}\widetilde{\xi}=R^{-1}\mfrac{\partial}{\partial x} \medint\int^{\widetilde{k\cdot x}}\!f(\widetilde{\xi}) \,\text{d}\widetilde{\xi}=n\, f(k\cd x) 
\end{gather}

There is no difficulty in extending these techniques to functions of more than one variable by introducing several rotation matrices,

\begin{gather}
\mfrac{\partial}{\partial x} \medint\int^{\widetilde{n\cdot x}}\!f(R_\text{1}\widetilde{\xi},R_\text{2}\widetilde{\xi},..)=n\,f(k_\text{1}\cd x,k_\text{2}\cd x,..)
\end{gather}

\bibliographystyle{unsrtnat}  
\bibliography{/home/hartin/Physics_Research/mypapers/hartin_bibliography}

\end{document}